\documentclass{article}
\usepackage{PRIMEarxiv}
\usepackage{amsmath}
\usepackage[T1]{fontenc}    
\usepackage{hyperref}       
\usepackage{url}            
\usepackage{booktabs}       
\usepackage{amsfonts}       
\usepackage{nicefrac}       
\usepackage{microtype}      
\usepackage{lipsum}
\usepackage{fancyhdr}       
\usepackage{graphicx}       
\usepackage{xcolor}

\title{The gravity model for social systems}

\author{
  Rafael Prieto-Curiel\\
  Complexity Science Hub \\ Metternichgasse 8 \\ 1030 Vienna, Austria \\
  \texttt{prieto-curiel@csh.ac.at} \\
}

\begin{document}
\maketitle

\vspace{30pt}

\section{Abstract}

Gravity is one of the most prominent models used across various social areas, including economics, demography, mobility, politics, and other systems where spatial interactions are relevant. The model represents a flexible approach that captures important regularities that might be detected when thousands or millions of people are observed, i.e.\ bigger origins and bigger destinations increase the intensity of interactions, but it tends to decay with longer distances between them. Therefore, the number of interactions between two locations is frequently modelled with an equation that resembles Newton's Law of Gravitation (also known as the Law of Gravity). Here, I explore different aspects, i.e.\ essential components, techniques for estimating the parameters, interpretation of results, and settings where gravity is a helpful tool, including mobility and migration.

\vspace{30pt}

\section{Introduction}

\begin{figure}
\centering

\includegraphics[width=0.3\textwidth]{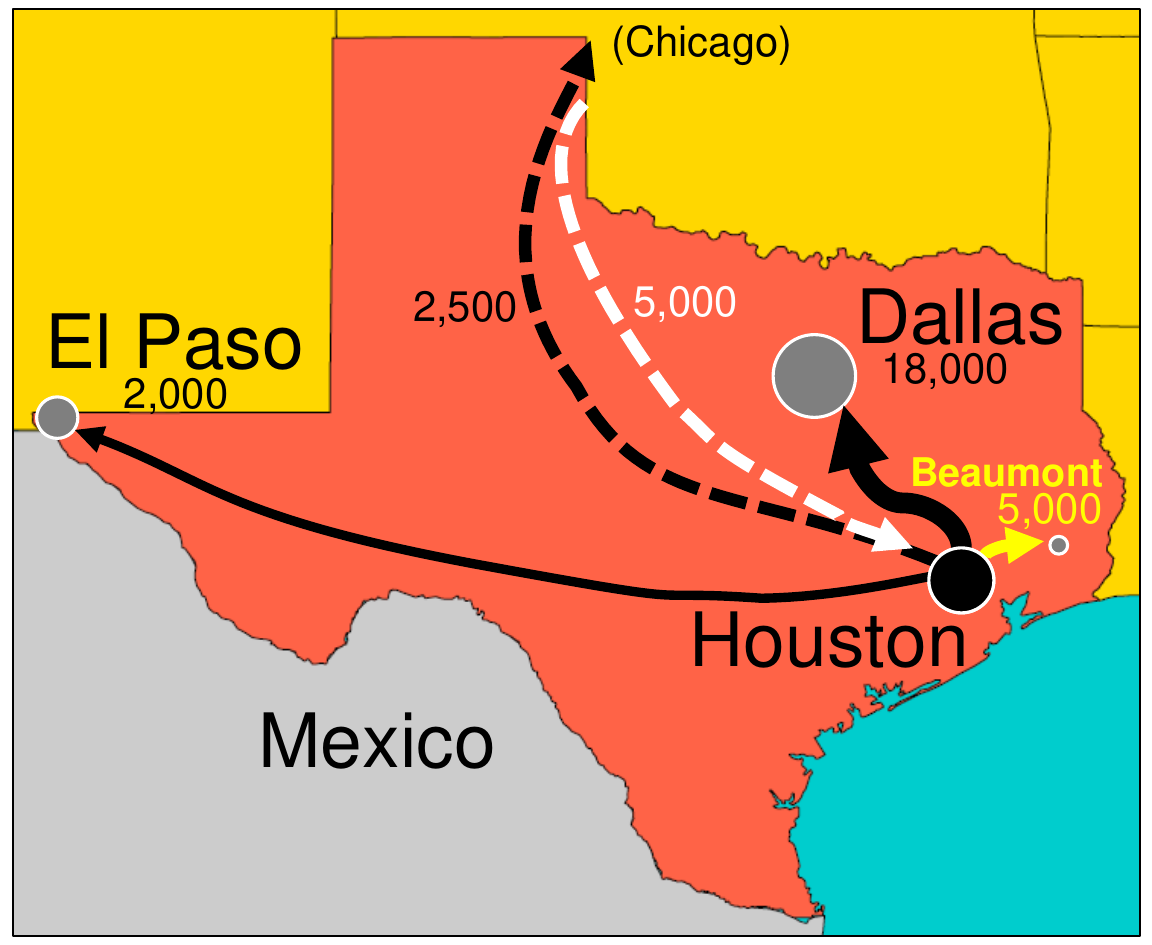}
\caption{Migrants to and from Houston, Texas, in 2018. The width of the arrows represents the intensity of the flow.}
\label{gravity:houston}
\end{figure}

Let us start with a real case of migration. Consider Houston, Dallas and El Paso, three cities in Texas, USA. In 2018, more than 18,000 people moved from Houston to Dallas, but only 2,000 moved from Houston to El Paso \footnote{Data from \url{https://www.census.gov/data/tables/2018/demo/geographic-mobility/metro-to-metro-migration.html} Last accessed in June 2024.}. Why is Dallas a more common destination for people from Houston than El Paso? There are too many reasons why people move and choose, among many possible destinations, one city as their new home. It is impossible to model all the reasons behind moving at an individual level. We can detect a collective pattern when we observe thousands or perhaps millions of people deciding whether or not to move and where to move. In particular, Dallas is only 384\,km away from Houston, whereas El Paso is more than 1,000\,km away. Dallas has nearly nine times more population than El Paso. If a person moves for work, to study, or to meet a family member, it is more likely that the person will find a job, an offer at a school, or a family member in Dallas than in El Paso (Fig.~\ref{gravity:houston}). 

\vspace{30pt}

The size of the destination and the physical distance are the main factors that help us model a social process in which geography plays a role. Perhaps at an individual level, people do not frequently think of the city size they are moving to, but rather, reasons that attract people, as many as they can be, are more commonly found in larger cities. But the size of the destination is not the only reason. From Houston, for example, nearly 5,000 people move each year to Chicago, a huge city (even larger than Dallas) in the Northern part of the US. The distance between Houston and Chicago is nearly 1,800\,km, so a person would have to drive more than 16 hours to return home (as opposed to Dallas, which is only 3 hours away). The likelihood of moving to a new destination decays with distance. Finally, suppose instead of considering people who moved \emph{from} Houston, we consider people who moved \emph{to} Houston. In that case, only 5,000 people moved from Beaumont (a medium-sized city one hour away from Houston) against the 16,000 people who moved from Dallas. As with the destination, we expect more people to move to Houston from larger cities simply by frequency.

Whether it is intracity migration, the number of visitors to a tourist destination, the number of journeys inside a city or the number of customers of a restaurant at the beach, gravity is one of the most powerful yet simple models to explore \cite{philbrik1973short}. Whether we apply it for mobility, migration, trade, or others, the gravity model\index{gravity model} for human constructs should be considered a general and very flexible model, which captures the impact of size at the origin and destination and their distance \cite{jung2008gravity}. It is a spatial model used for decades, perhaps under different names but capturing similar principles \cite{philbrik1973short, GravityModel, MigrationLaws1885, stouffer1940intervening}. The gravity model usually depends on the size of the origin, usually captured by its population. The flow of people, specific products, or the number of migrants is expressed as an increasing function of the size of origin, assuming that a higher flow is expected with more people. The gravity model captures the \emph{attractivity} of the destination and, in general, utilizes population size as a proxy. However, the size of the origin might be captured by different means. For example, we could use GDP if we are interested in trade or the number of hotel rooms for an application in tourism \cite{batty2021new}. In terms of the distance, the actual physical distance between origin and destination is frequently used, but the model allows other forms of pairwise relationships that might capture proximity, such as travel time or other aspects, including frictions or barriers \cite{Gravity, ZIPMigration, NorthAmericaMigration}. The gravity model, for example, has been used to model trade between countries, considering the population (or income) size and the physical or cultural distances between two distinct locations \cite{Gravity, GravityModel, barbosa2018human}.

Let $F_{o,d}$ be the number of ``visitors'' from $o$ to $d$ during some period. That can be the number of customers from the neighbourhood $o$ to restaurant $d$ or the number of people who moved from Houston to Dallas, so $F_{Houston, Dallas} = 18,000$. Then, we aim to construct a model such that $F_{o,d}$ increases with the size of the origin $P_o$, also increases with the size of the destination $P_d$ and decreases with the distance between them, $D_{o,d}$. Many equations satisfy those three principles. Commonly, one such expression is used, as it has many desirable properties, including parameters that we know how to interpret and ways to estimate them \cite{GravityR}. Such an equation is frequently expressed as
\begin{equation}\label{gravity:Gravs}
F_{o,d} = \kappa \frac{P_o^\mu P_d^\nu}{D_{o,d}^\gamma}
\, ,
\end{equation}
where $\kappa$, $\mu$, $\nu$, and $\gamma$ are non-negative model parameters that we frequently estimate with data or make some ``educated guess'' based on some knowledge of the process. $\mu$ and $\nu$ capture non-linear impacts of the size of origin and destination; $\gamma$ captures the impact of distance, and finally, $\kappa$ is a constant which helps adjust figures and units. For some values of $\kappa$, $\mu$, $\nu$ and $\gamma$, the Eq.~(\ref{gravity:Gravs}) may be interpreted as the expected frequency at which people from the $n$ origins visit the $m$ destinations. 

The expression, with $\mu = \nu = 1$, and with $\gamma = 2$, was conceived by Isaac Newton in 1687 for modelling the attraction between two objects, and it is usually called ``the law of universal gravitation''. Due to its resemblance to the law of universal gravitation, the social model is called the \emph{gravity model}. The gravity model in social settings has many expressions, one of which mimics Newton's law of universal gravitation. However, many expressions capture similar aspects, for instance, using exponentials instead of functions like $P_o^\mu$ or $P_d^\nu$.

Newton noticed that to transform the product of two masses divided by the square of their distance from the right-hand side (with units $kg^2 / m^2$) to a force on the left-hand side (with units $kg m/s^2$), he needed a constant with funky units, known as the \emph{gravitational constant} (approximately $6.674\times 10^{-11} m^3/kg s^2$. The Cavendish experiment measured the gravitational constant more than 100 years after the law of universal gravitation was published. In our expression, $\kappa$ acts as such a constant, which helps transform between units. The gravity model is frequently used to model the flux for a given period, usually one year in the case of migration and one day or week in the case of visitors. In the model, the value of $\kappa$ depends on the interval of time considered. If we double the time interval, we should also double the values of $\kappa$ to compensate. 

The gravity model for human constructs should be thought of as a general and very flexible model, which captures the impact of size and distance and might capture other aspects, such as frictions or barriers \cite{Gravity, GravityModel, ZIPMigration, NorthAmericaMigration}. Different expressions of the gravity model are frequently used to inform relevant aspects of migration, transport, mobility or trade, so they are a valuable tool for policymakers \cite{OECDMigration, RootExodus}.

The model should not be conceived as Newton's gravity equation with one or two parameters to fit. Migration between two cities does not double if the population of the destination doubles, and migration between two cities is not four times larger with half the distance. Nothing forces humans to behave in such strict manners. Two particles are attracted equally under similar circumstances, but no two individuals are so identical to ensure they will visit or move to the same destination under the same circumstances. Thus, any physical model used in a social setting must be carefully analyzed, including all assumptions needed to use a specific expression. Only with one particular expression, and if we fix the parameters (with $\mu = \nu = 1$ and $\gamma=2$), we obtain Newton's universal gravitation law.

The gravity model captures the \emph{attractivity} of the destination and utilizes size as a proxy. However, other metrics may be used, for instance, the number of beds in a tourist destination or job offers. It depends on the distance, but it can be physical distance, road distance, monetary costs, and time travel. The model has some parameters (perhaps too many) and at least four. The gravity model is very flexible, and many insights may be obtained from it \cite{philbrik1973short}. With the correct data, it is possible to detect if physical distance has a higher impact among young people or the elderly or to check if people prefer moving to locations with similar weather. Using a gravity-based model, it was detected that people prefer moving to cities of similar size as their origin, and rarely does a person from a big city move to a small town or the countryside, and vice-versa \cite{ScalingMigrationRPC}. If a company is deciding where in the city to open a new store, they can estimate how many people would visit it from different neighbourhoods. Also, using some of the principles and techniques from the gravity model, they can estimate how many customers they could lose at their other locations. The gravity model depends on data and might rely on some assumptions that must be carefully considered. There are many techniques to estimate its parameters, ranging from simple linear regression to minimizing the differences between what is observed and modelled.

\section{The building blocks of gravity}
Whether we are using the gravity model to forecast or detect the impact of some environmental or social factors, the first step is to define the four components clearly: origins, destinations, distances and fluxes.
\begin{itemize}
\item \emph{Origins} -- define if the model is based on cities, metropolitan areas, countries, neighbourhoods, or similar spatial units and clearly define what we mean about size. Frequently, the population size of each origin is the most suitable metric.
\item \emph{Destinations} -- perhaps the easy choice here is to mimic origins, so consider flows between cities or trade between countries and use their corresponding size. Perhaps the model considers tourist destinations or shopping malls, and the corresponding size metric is the number of hotel rooms or the surface of distinct commercial areas. Depending on the considered indicator, the destinations and the corresponding way they are measured might be modelled in different ways. It quantifies how attractive each destination is. 
\item \emph{Distances} -- depending on the objective, the distance might be commuting time, actual physical distance, costs or other units. Often, the objective of gravity models is to detect the impact of barriers, like an international border, certain interventions, like a new train station, or some social factors, such as the impact that cultural or language differences impose on migration. 
\item \emph{Flux} -- some data which captures the phenomenon of the model is needed. Suppose the objective is to capture tourism, for instance. In that case, some data that suits the origins and destinations is needed, such as the number of visitors from different origins to each destination during a certain period. A frequent issue with this type of data is having a very short period for a process that takes longer times or dealing with missing or incomplete observations. 
\end{itemize}

For the notation, usually, for origin $o$ with size $P_o$ and destination $d$ with size or attractiveness $P_d$, the flux (or number of migrants, say) between $o$ and $d$ is written as $F_{o,d}$ and the distance is $D_{o,d}$. When the flux is estimated, usually we write $\hat{F}_{o,d}$. Also, it is common to think of the total outflow of some origin (the sum of $F_{o,d}$ over all possible destinations), and it is usually expressed as $\sum_d F_{o,d} = F_{o, \bullet}$. Similarly, the total inflow is the sum over all possible origins, and it is expressed as $\sum_o F_{o,d} = F_{\bullet, d}$.

\subsection{Assumptions}
All models rely on data and assumptions. In the case of the gravity model, it is worth considering first if a linear impact of the origin and destination sizes $P_o$ and $P_d$ is reasonable. Many things have been detected to scale non-linearly with city size, for instance, GDP \cite{GrowthBettencourt}. That means that if city B is ten times larger than city A, then the GDP of city B is substantially more than ten times the GDP of city A, which means that city B has a higher GDP per person than city A. Migration between cities is affected by city size, meaning that assuming a linear model might be too restrictive \cite{ScalingMigrationRPC}. However, assuming a linear impact of the origin size might be reasonable for other aspects. For example, if we deal with some trade data, thinking that a city with twice the number of people consumes two times the amount of food or twice the number of beers might be reasonable. Still, it is worth checking if some non-linear departures with size are observed, and often, the linearity assumption is too restrictive.

A second aspect to consider is symmetry. The gravity model is often applied to a set of units, for example, migration between cities or trade between countries. Therefore, the gravity model frequently considers some units (say, a city) as the origin and destination of some flux. Then, the flux and the distances can be structured on a square matrix, where rows and columns correspond to separate units. Notice that distances are symmetric (the matrix $D_{o,d}$), and so the right-hand side of Eq.~(\ref{gravity:Gravs}) gives the same values if $\mu = \nu$. Unless the parameters $\mu$ and $\nu$ in Eq.~(\ref{gravity:Gravs}) have different values, symmetry is somehow being assumed and forced into the model. Thus, it is assumed that outflows and inflows are equal. Unless there is a solid reason to think that the process is symmetric, it is preferable to let $\mu$ and $\nu$ as different parameters (and perhaps obtain symmetry as a result of the parameter estimation, if $\hat{\mu} \approx \hat{\nu}$).

\subsection{Measuring units}
One issue with the gravity equation is related to the units. The product of the population of two cities on Eq.~(\ref{gravity:Gravs}) could have too many digits, and divided by some distance in km, it still has too many digits. Therefore, often, to get an estimated flow (which frequently is not that big), the estimated value of $\hat{\kappa}$ in Eq.~(\ref{gravity:Gravs}) has to be very small to compensate. Dealing simultaneously with huge and tiny numbers may lead to numerical errors, so the best thing to do is to think first of the units. Depending on the case, expressing population size in millions and distance in hundreds of km might prevent dealing with large and small units simultaneously.

If the distance between two observations is very small, then the impact that distance has on Eq.~(\ref{gravity:Gravs}) might not be correct and might lead to an unbounded estimate. In such cases, perhaps it is worth considering those two units as the same and merging the number of visitors from them.

Something to consider concerning the gravity model is that it requires large amounts of data, including population sizes, distances and the flow between every pair of cities. For instance, the gravity model for the $N = 118$ cities with more than 100,000 inhabitants in Nigeria requires capturing 118 population sizes, measuring 7,000 distances $D_{o,j}$ and knowing roughly $N(N-1) \approx 14,000$ flows $F_{o,d}$.

\section{Applying the gravity model}
\begin{figure}
\centering

\includegraphics[width=0.6\textwidth]{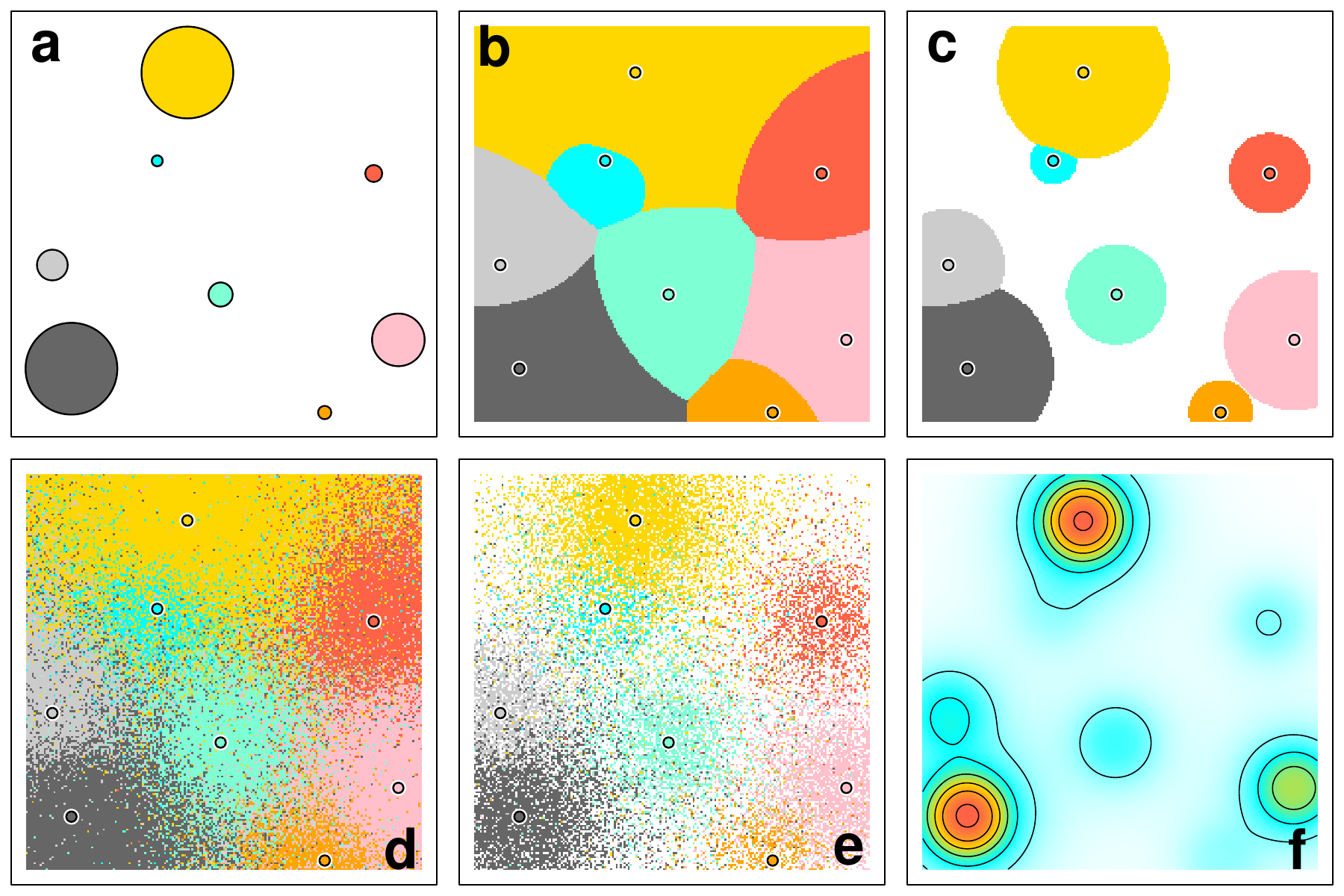}
\caption{(a) Different ways in which the attractiveness to some place can be observed, where the location of $m = 8$ destinations is shown by the size of the disc. (b) Detection of the most attractive destination for each square in the grid.  (c) Detection of the most attractive destinations, where the attractiveness is larger than some positive threshold, so the attractiveness below some threshold is not coloured. (d) Selection of a destination based on the probability of being selected is proportional to the attractiveness of each destination. (e) Selection of a destination based on the probability of being selected is proportional to the attractiveness of each destination, considering also the probability of not moving. (f) Total gravity is computed as the sum of the attractiveness of each destination experienced by each square in the grid (f).}
\label{gravity:SixFigures}
\end{figure}

Consider, for example, a $n \times n$ square grid where one person occupies each square. On top of that grid are $m$ destinations (say, stores or cities) with some level of attractiveness $P_k$ for each destination (with $k = 1, 2, \dots, m$). Person $i$ is attracted to destination $k$ according to
\begin{equation}
\label{gravity:GravsApp}\nonumber
A_{i,k} = \frac{P_k^\nu}{D_{i,k}^\gamma}
\, ,
\end{equation}
where $D_{i,k}$ is the Euclidean distance between the square $i$ and the destination $k$ and $\nu>0$ and $\gamma>0$ are some parameters. Here, each square has some ``gravity'' to each destination. For some values of $\nu$ and $\gamma$, it is possible to detect, for each person, what is their most attractive destination. The most attractive destination for most people is frequently the nearest one, but it depends strongly on attractiveness (Fig.~\ref{gravity:SixFigures}a). For example, if we consider cities in some countries instead of destinations, their attractiveness could be their size or the number of jobs available. If each person goes to their most attractive destination, the result is partitioning the plane into regions (Fig.~\ref{gravity:SixFigures}b).

With a gravity model, it is relevant to consider that choosing the most attractive destination is always possible, no matter how far or attractive the destinations are. Therefore, if we consider that the destinations are shops where people might move, we might assume everyone picks a shop. However, this is not necessarily the case. People far away from all destinations might choose not to go to any shop (Fig.~\ref{gravity:SixFigures}c). An option here is to consider the attractiveness of any destination above a certain threshold. That is, take some value of $\omega > 0$ and consider only values of the attractiveness where $A_{i,k} > \omega$.

The gravity model returns the attractiveness from each location to every destination. The level of attractiveness of some destinations can also be used to consider the probability that a person will select them. For example, let's ask a person to choose among their $m$ options. The probability that they choose destination $k$ is given by $\pi_{i,k} = A_{i,k}/\sum_k A_{i,k}$, so the sum of the probability of selecting all destinations is one. This gives us a stochastic model, where the probability of moving to some destination is a realization of a multinomial distribution (Fig.~\ref{gravity:SixFigures}d). The term ``multinomial'' here refers to a probability distribution that can be thought of as if the person rolls an ``unfair'' dice with $k$ sides and picks the corresponding destination (unfair since the probability of each side will not be the same). Again, if a person is too far from any destination, they might choose not to move. The model can be easily adjusted if we consider an additional ``destination'' that represents the option of not moving (Fig.~\ref{gravity:SixFigures}e). 

For each square in the grid, we can consider the maximum attractiveness they receive from all possible destinations or the total attractiveness (Fig.~\ref{gravity:SixFigures}f) computed as the maximum attractiveness or the summation of all destinations. These types of computations might be useful to detect accessibility \cite{guagliardo2004spatial}. Also, these calculations help plan strategies to detect where to open a new store, for example, by adding a new destination to the data and measuring how many customers they expect. A similar model was constructed nearly a century ago to detect the ``point of indifference'' between two retail centres, using the size of each destination and the distance \cite{reillylaw}.

\subsection{Impact of the parameters}
\begin{figure}
\centering

\includegraphics[width=0.5\textwidth]{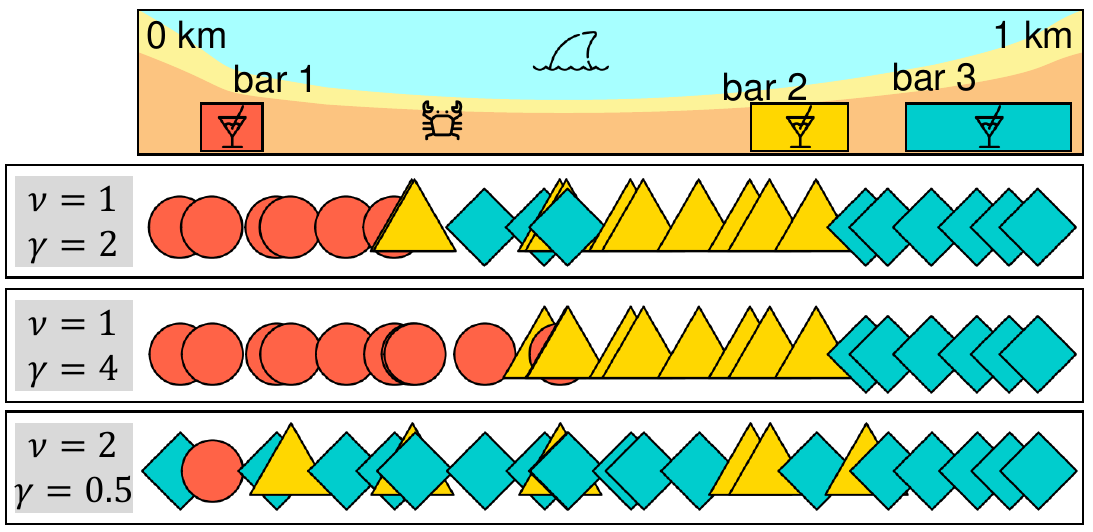}
\caption{Three bars located on a 1\,km beach. Bar 1 is located 100\,m away from one beach edge, Bar 2 at 700\,m and Bar 3 at 900\,m. The corresponding size of the dance floor at each bar is 5, 10 and 20 m$^2$. At the beach, there are 25 people located in random positions. Each person chooses a bar based on their position and the size of the dance floor. They prefer bars with a bigger dance floor, but they would also rather walk short distances at the beach. Each customer is represented by a red circle when they choose Bar 1, a yellow triangle when they choose Bar 2, or a blue square when they choose Bar 3. Different values of $\nu$ and $\gamma$ in Eq.~\ref{gravity:Gravs} change how bars are picked. For instance, for large values of $\gamma$, the size becomes less relevant, and it's ``all about location''. For larger values of $\nu$, the size of the dance floor becomes more relevant, so bigger bars are more popular.}
\label{gravity:ApplyGravity}
\end{figure}

We can see the impact of the parameters with a simple example. Imagine a 1\,km long beach that has only three bars. Every day, $n$ people go to the beach and decide which bar to visit. The size of the dance floor of each bar is what makes them more attractive, so consider that their sizes are $P_1 = 5$, $P_2 = 10$ and $P_3 = 20$ m$^2$, located at 100, 700 and 900\,m from one of the edges of the beach (Fig.~\ref{gravity:ApplyGravity}). The probability that the person $i$ goes to bar $k$, written as $\pi_{i,k}$, is proportional to 
\begin{equation}\label{gravity:SingleAtt}
\pi_{i,k} \propto \frac{P_k^\nu}{D_{i,k}^\gamma}
\, ,
\end{equation}
where $D_{i,k}$ is the distance between the person $i$ and the bar $k$. The symbol $\propto$ means ``proportional'', and it is frequently used to simplify the notation. Remember that we deal with probabilities, so $\sum_{k = 1,2,3} \pi_{i,k} = 1$. Thus, the bar that each person could choose can be modelled with a multinomial distribution, and the bar they go to is a realization of that distribution. 

Simulating how a customer picks a bar at the beach is possible. Bars are visited more frequently by nearby customers. However, for higher values of $\nu$, people care more for the size of the dance floor, whereas higher values of $\gamma$ mean that distances are more relevant. This version of the gravity model is known as the \emph{Huff model}, formulated in 1963 by David Huff \cite{HuffModel, luo2014integrating}. It is frequently used in economics, retail research, and urban planning to capture ``competition'' between distinct destinations, such as bars. 

If bars are too small or far away, people might not visit them! In the Huff model, every customer has to choose a bar, regardless of how far away or unattractive they might find them. But what if $P_k^\nu / D_{i,k}^\gamma$ is too small for all bars? It is possible to extend the Huff model so that not everyone has to pick a bar.

\subsection{Gravity model on a network}
Thinking of the gravity model on a grid is the simplest one since each person is represented by a square in the grid (so each square is one person). What if, instead of one person on each square, we imagine there are $P_i$ people on each square? For example, if each square was a neighbourhood in a city and the $m$ destinations were shops where people would buy groceries. Instead of an entire grid, we can imagine only a few squares (or origins) with $P_i$ people on each, with $i = 1, 2, \dots, n$. So, we consider $n$ origins with $m$ destinations. The origins and the destinations do not need to be the same (although, in some applications, it makes sense to consider the same set). For example, people across $n$ different neighbourhoods (each neighbourhood with $P_i$ population) might shop in $m$ stores (each store with some attribute $S_k$ that is the attractiveness of each store). This network can either be a bipartite with $n$ nodes as the origins and $m$ nodes as the destinations or a network with $n$ nodes if origins and destinations are the same set. 

In this case, we assume that people of each origin act similarly, so they have the same probability of going to any of the $m$ stores or moving to any of the $m$ cities. Instead of looking at probabilities in Eq.~(\ref{gravity:SingleAtt}), we look at frequencies when we multiply both sides with the population of the origin and obtain $F_{o,d} = \kappa \frac{P_o^\mu P_d^\nu}{D_{o,d}^\gamma}$, i.e.\ Eq.~(\ref{gravity:Gravs}).

Suppose we have $n$ origins (that represent neighbourhoods in a city) with $m$ destinations (for example, green areas in the city). For each origin, we consider its population $P_o$, and for each destination, we consider its level of attractiveness, $P_d$. We then measure the distance between each origin and destination, $D_{o,d}$, which might be the Euclidean distance, but also the road distance, commuting times or other units that reflect proximity \cite{prieto2021detecting}. For example, some distances could be zero when the origins and destinations are the same. However, the gravity model is designed to capture the impact of \emph{long} distances. Therefore, avoiding distances equal to zero (or even too close to zero) is better.

\section{Using the gravity model}
Although Eq.~(\ref{gravity:Gravs}) may be used to estimate the flow of people or trade between distinct locations, the model is frequently used to measure the impact of distance or political barriers. Based on some data (that is, having values of $F_{o,d}$ for different origins and destinations and having also defined the size of the origin, the attraction of the destinations and the distance between them), a common challenge is to estimate the values of the parameters. The gravity Eq.~(\ref{gravity:Gravs}) has at least two parameters, $\kappa$ and $\gamma$ if a linear and symmetric impact of city size was assumed (so that $\hat{\mu}=\hat{\nu}=1$ as an assumption) or three or four parameters, depending on how it was conceived. 

Although rarely observed, one ideal scenario is when we have a complete dataset (that is, no $F_{o,d}$ is missing) and all flux values are well above zero. Then, we can take the logarithm on both sides of Eq.~(\ref{gravity:Gravs}), and the result is
\begin{equation}\label{gravity:Logs}
\log F_{o,d} = \log \kappa + \mu \log P_o + \nu \log P_d + \gamma \log D_{o,d}
\, .
\end{equation}
The expression is a linear equation concerning the logarithm of the populations and distances. The sign in front of $\gamma \log D_{o,d}$ should be negative, but we can also write it with a positive sign and keep in mind that $\gamma$ should be negative. Also, $\log \kappa$ can be considered a new constant number. Then, the coefficients $\kappa' = \log \kappa$, $\mu$, $\nu$ and $\gamma$ need to be estimated. Assuming that our model gives an unbiased flow estimate, a regression gives us the four parameters. If we further assume that everything else that is not captured only by population size and distance follows a normal distribution with constant variance (in  Eq.~(\ref{gravity:Logs}), with logarithm on both sides). In that case, we also get estimates of the variance of each parameter so that we can also obtain statistical significance. Thus, parameters may be obtained with a regression. 

Almost always, there will be pairs of cities for which the number of visitors is zero, meaning that there will be some $F_{o,d}=0$ for which the logarithm is not defined. Adding some small constant $\epsilon$ to avoid zeros might create substantially different estimates, depending on how small or large is that arbitrary $\epsilon$, so it might create more issues and uncertainty \cite{ZIPMigration}. A better option might be to use a Poisson regression, which is appropriate when the dependent variable is a count, such as the number of people, transactions, or others. In a Poisson regression, the flow $F_{o,d}$ is assumed to follow a Poisson distribution with rate $\exp(\kappa +\mu P_o + \nu P_d + \gamma D_{o,d})$ and the estimation works similarly with a regression. 

Here is where other elements might be added to the model. For example, the unemployment rate at the city of origin and destination might help capture its impact on pushing and pulling people. We could also consider a binary variable $I_{o,d}$ that has values of one if two cities share some attribute (say, they are in the same state or have similar weather) and zero otherwise. See \cite{GravityModel, GravityPanelData, NorthAmericaMigration} for ways in which parameters might be estimated, and see \cite{ZIPMigration} for the challenges that the estimation poses. 

Besides regression techniques, the gravity model can also be considered an optimization problem. For a set of values $(\kappa, \mu, \nu, \gamma)$, the right-hand side of Eq.~(\ref{gravity:Gravs}) produces some estimate of the flow $\hat{F}_{o,d}$ and some error $F_{o,d} - \hat{F}_{o,d}$. Adding all the square values of the errors gives a function of the four parameters, which then should be minimized.

\subsection{Parameter interpretation}
Even before estimating the parameters, it is worth thinking about the expected values of each parameter and letting the regression or the optimization only confirm our impressions. The coefficients $\mu$ and $\nu$ affect city size. Suppose the smallest city among our observations has $P_s = 100,000$ inhabitants and the largest has $P_l =10$ million so that $P_l = 100 P_s$. With a value of $\hat{\mu} = 1$, the model says that the probability of moving is the same for the two cities. That is, the flow is proportional to city size, so we expect 100 times more flow from the large city simply because it is 100 times larger. However, with $\hat{\mu} = 1.15$, we already get that the flow for the large city is $P_l^{1.15} = (100 P_s)^{1.15} \approx 200 (P_s^{1.15})$ so that the flow with respect to population is twice as intense as it is for the small city. Things might explode quickly for these types of coefficients. With $\hat{\mu} = 1.5$, the process is ten times more intense (or the frequency of migration is ten times larger for the biggest city), and with $\hat{\mu}= 2$, it is 100 times more intense. In the case of $\hat{\mu} = 2$ and if 1\,\% of the population from the small city moves, then 100\,\% of the population of the large city moved, so the value is already outside a reasonable range. We get a similar effect for values smaller than one, where values below $\hat{\mu} = 0.05$ should not be observed. And the same occurs with $\hat{\nu}$. Values close to 1 are reasonable, and above 2 or below 0.05 are not. For values of $\hat{\mu}>1$, the process is more intense in large cities, and for values of $\hat{\mu}<1$, the process is more intense in small cities. If uncertain, consider what happens on both sides of Eq.~(\ref{gravity:Gravs}) when you divide by the size of the origin (or the destination). The left-hand side gives the fraction of visitors, and the right-hand provides some function of the sizes and distances.

The relevant null hypothesis to test concerning $\mu$ and $\nu$ is whether their estimated values might equal one. If an interval for $\hat{\mu} \in [a,b]$ contains 1, there is not enough evidence to say that visitors are more intense in larger or smaller cities.

For distance (whether physical, road distance, commuting time or other distance measure), a value of $\hat{\gamma}$ close to zero means that it is not very significant, and higher values indicate that the observed process is highly sensitive to distance. With $\hat{\gamma}=1$, the estimated flow between two cities decreases by half if the distance between two cities doubles. With $\hat{\gamma}=1/2$, if the distance doubles, the estimated flow is reduced only by 30\,\%.

In general, a regression (or an optimization problem, minimizing the mean square error) is highly sensitive to extreme values, so if the data considers two very close or distant cities, those values will greatly impact the parameters. The same happens with the size of origins and destinations. If the model has a huge or tiny city, many results will be based on the corresponding flow. And the same might occur with migration or the number of visitors. If, for some reason, there is a very high flow between two cities (for instance, if a large company decides to move its offices one year), then that process will have an enormous impact on the estimation process. 

\begin{figure}
\centering

\includegraphics[width=0.5\textwidth]{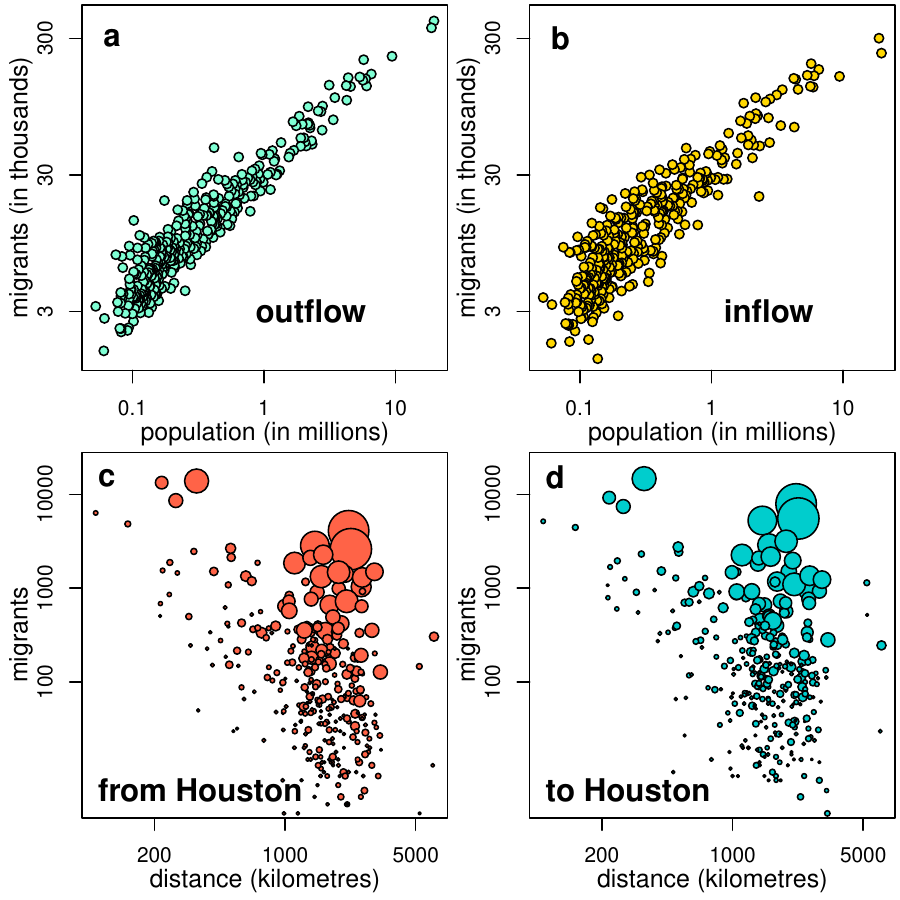}
\caption{(a) Number of outflow and (b) inflow migrants from one year to the next one between US metropolitan areas in the vertical axis and the corresponding population in the horizontal axis. (c) Number of migrants from and (d) to Houston (bottom) from different metropolitan areas (vertical axis) and the Euclidean distance between Houston and other areas (horizontal axis). The size of the disc is proportional to the population of the city. All panels are on a logarithmic scale on both axis. }
\label{gravity:USFigures}
\end{figure}

A useful tip that works here but is also valuable in other social areas is plotting the data and checking if the model and the assumptions make sense (Fig.~\ref{gravity:USFigures}). For the gravity model, four very useful plots are: 
\begin{itemize}
\item size of the origin $P_o$ in the horizontal axis and the total outflow $F_{o, \bullet}$ in the vertical axis,
\item size of the destination $P_d$ in the horizontal axis and the total inflow $F_{\bullet, d}$ in the vertical axis,
\item for one observation (perhaps a large city or a small one), take the distance to other observations in the horizontal axis and the outflow from it in the vertical axis, and
\item for the same observations, consider the inflow to it in the vertical axis. 
\end{itemize}

The idea with these four plots is to observe a higher total inflow and outflow from bigger cities. We expect longer distances to discourage interactions with the third and fourth plots so there are fewer migrants in both directions. Here, we observe the impacts of the distance between cities and their size (Fig.~\ref{gravity:USFigures}, bottom). There is a general trend where longer distances discourage migration, and more populated cities (so bigger discs) attract and push more people from and to Houston. Any parameter fitting should come after we are sure the data is correct and the model's assumptions correspond with the observed patterns.

\subsection{City to city migration}
We can use the gravity model to verify some aspects of trade between countries or to model migration. For instance, when people move between cities in the US, do they prefer to stay in the same state? The American Community Survey captures metro area-to-metro area migration flows, so it helps us answer this question. Let $F_{o,d}$ be the number of people that moved from some city $o$ to some destination $d$ in a year, and let $P_o$ and $P_d$ be the corresponding populations (in millions). Also, let $D_{o,d}$ be the Euclidean distance between cities (in km). Taking the logarithm on both sides of Eq.~(\ref{gravity:Gravs}), we obtain that
\begin{equation}\label{gravity:MigL}
\log F_{o,d} = \log \kappa + \mu \log P_o + \nu \log P_d + \gamma \log D_{o,d}
\, .
\end{equation}
Theoretically, we could adjust a regression and obtain values of $\kappa$, $\mu$, $\nu$, and $\gamma$. However, is it possible to take the logarithm on both sides? Not really. Of all the possible pairs of cities in the US, there are zero migrants in 67\,\% of them, so the logarithm on both sides of the equation is not well defined. Thus, it is better to use a Poisson regression here. Instead of Eq.~(\ref{gravity:MigL}), we write
\begin{equation}\label{gravity:MigN}
E [F_{o,d}] = \exp \left( \log \kappa + \mu \log P_o + \nu \log P_d + \gamma \log D_{o,d} \right),
\end{equation}
where the left-hand side of the equation is the expected number of migrants from $o$ to $d$, given the corresponding sizes and distances. The sign of $\mu$ and $\nu$ should be positive, and we expect that distance discourages migration, so the sign of $\gamma$ should be negative.

Since we are wondering whether being part of the same state impacts the intracity migration propensity, we define the binary variable $S_{o,d} = 1$ if both cities belong to the same state and zero if they belong to a different state. We can add that variable to the regression and express it as
\begin{equation}\label{gravity:MigS}
E [F_{o,d}] = \exp \left( \log \kappa + \mu \log P_o + \nu \log P_d + \gamma \log D_{o,d} + \eta S_{o,d}\right),
\end{equation}
where $\eta$ is the impact of being in the same state. The results of the Poisson regression models are in Tab.~\ref{gravity:TableModels}. Model~1 only considers the size of the origin, the destination and the distance, whereas Model~2 also considers whether two cities belong to the same state.

\begin{table}
\caption{Results of the Poisson regression models. The dependent variable is the expected number of migrants between the two cities. Model~1 gives the values of the parameters of Eq.~(\ref{gravity:MigN}). Model~2 also considers the impact of two cities being part of the same state in the US, from Eq.~(\ref{gravity:MigS}).}
\label{gravity:TableModels}
\begin{center}
\begin{tabular}{l c c}
\hline
 & Model~1 & Model~2 \\
\hline
(Intercept) & $12.314 \; (0.002)^{***}$ & $10.036 \; (0.003)^{***}$ \\
$\log P_o$ & $0.870 \; (0.000)^{***}$ & $0.878 \; (0.000)^{***}$ \\
$\log P_d$ & $0.814 \; (0.000)^{***}$ & $0.821 \; (0.000)^{***}$ \\
$\log D_{o,d}$ & $-1.086 \; (0.000)^{***}$ & $-0.777 \; (0.000)^{***}$ \\
same\_state & & $1.348 \; (0.001)^{***}$ \\
\hline
AIC & $11,833,941$ & $9,979,479$ \\
Num. obs. & $147,840$ & $147,840$ \\
\hline
\multicolumn{3}{l}{\scriptsize{$^{***}p<0.001$; $^{**}p<0.01$; $^{*}p<0.05$}}
\end{tabular}
\end{center}
\end{table}

It is easy to interpret the results in Tab.~\ref{gravity:TableModels}. For example, for Model~1, we take two cities (their population in millions and the distance between them in km), and we estimate that the number of migrants between them is given by $E_{o,d} = \exp(12.314) \times P_o^{0.870} \times P_d^{0.814} \times D_{o,d}^{-1.086}$. Thus, we confirm that with a bigger origin and a bigger destination, we expect more migrants. The flow, however, is not linear, not symmetric and more intense in smaller cities (since both $\hat{\mu}<1$ and $\hat{\nu}<1$). Also (because of the negative sign), longer distances mean less migration. The impact of distance is quite significant. If, for example, the distance between two cities doubles, then the size of the destination needs to increase by 2.5 times to compensate (so our estimate remains constant). The distance coefficient is $-1.086$, meaning the impact is roughly linear, not quadratic. Nothing forces humans to behave like planets attracting each other.

For Model~2, we added a binary variable to check whether two cities in the same state have higher migration flows. The Akaike Information Criterion (AIC) evaluates how well a model fits the data. It is computed as $AIC = 2 k - 2 \log L$, where $k$ is the number of variables used and $L$ is the likelihood estimate. The AIC balances the model performance and the number of parameters used. Models with a smaller AIC are better, so adding more variables (larger $k$) is preferred only if it increases the likelihood sufficiently. We see that Model~2 is an improvement over Model~1.

\begin{figure}
\centering

\includegraphics[width=0.65\textwidth]{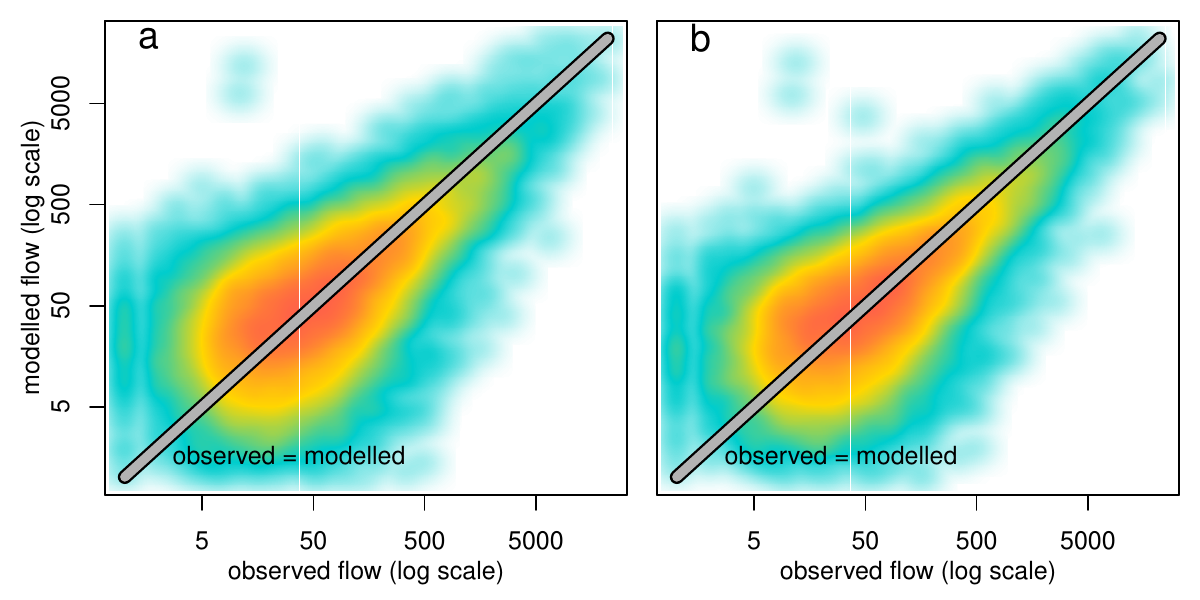}
\caption{Observed number of the metro to metro migrants in the US in a year between 2015 and 2019 (horizontal axis) and modelled number (vertical axis). Model~1 (a) only considers the impact of the origin and destination size, and Model~2 (b) considers whether two cities belong to the same state. This type of figure is a smooth scatterplot, which is useful for detecting patterns when we have thousands of data points. The part of the figure with more points is red, with fewer points are yellow, then blue, and a white region means no points. The dark diagonal line marks when the observed values match the predicted values for any two cities. The section above the line indicates the region where the model overestimates migration, and below the line, migration is underestimated. Both panels are in logarithmic scale in both axis.}
\label{gravity:Predicted}
\end{figure}

The results show that adding the binary variable $S_{o,d}$ to our model made it more accurate. Thus, there is an impact of two cities being from the same state and due to the sign of the variable, being from the same state significantly increases the migration between cities. However, the coefficient of $\log distance$ in Model~2 is smaller than in Model~1, reflecting perhaps that cities from the same state tend to be nearby.

How precise is the gravity model? One way to verify its precision is by plotting the observed flow (so the number of migrants between every pair of cities in our case) against the predicted number (Fig.~\ref{gravity:Predicted}). Since the size of cities covers a wide range and the intracity migration also has a wide range of values, using a logarithmic scale on both axes is common. Values closer to the diagonal mean the predicted values are close to the observed values. For a small number of visitors or migrants, the gravity model tends to give values that may be perceived as far from the observed figures, but it means that the gravity model shows values that are maybe a few dozen away. When there are many migrants between two cities, the gravity model seems more accurate. And the gain of adding the binary variable $S_{o,d}$ is that the predicted values, particularly the large ones, are closer to the observed values.

\section{Conclusion}
Predicting whether a person will move to some city, travel to some neighbourhood, or visit a park or a shop is very challenging at an individual level. But when we observe hundreds or maybe thousands of people, we can detect patterns. Based mainly on the origin's size, how attractive the destination is, and the distance between them, the gravity model is designed to capture most aspects of that collective pattern.

The gravity model is frequently used when data is scarce to obtain some rough estimate of the flow between two locations, simply applying Eq.~(\ref{gravity:Gravs}) or some variation directly to the data. For each person, we are considering the process of choosing among many options as a multinomial variable. Also, repeating the same simulations makes it possible to obtain intervals, so the model is quite powerful in detecting departures. With enough data, the gravity model may also be used to fit a regression and fetch values of the parameters. With those parameters, we can infer what attracts people to some destination, including geographic, demographic, economic, climatic, and environmental factors \cite{garcia2015modeling}. 

Some caution is needed when using the gravity model. Many assumptions are involved with the gravity equation, and often, these assumptions are not desired. For example, in Eq.~(\ref{gravity:Gravs}), if we assume that $\mu = \nu$, we impose symmetry on the flow, which is often not the case. Applying the gravity model without considering all assumptions and parameters could yield wrong results. For example, applying the gravity model using some temporal data might give some misleading forecasts \cite{beyer2022gravity}. The model is not bounded, and in some scenarios, it might also predict that more people will leave the origin than there are \cite{BarabasiMigration}. Also, two numerical considerations are relevant for the computation of the gravity model. Some of the values may be simultaneously too large and too small. For example, when two populations are multiplied in Eq.~(\ref{gravity:Gravs}). Avoiding very small and very large units in the same equation gives better numerical estimates. 

One of the most important things to remember about the gravity model is that some physical models might help us structure a social process systematically and concisely. They work great to form an analogy between known processes and their equations and social behaviours. However, humans do not behave like planets in the solar system or like drops of water, so no model will perfectly capture all aspects of migration.

\section*{Acknowledgements}

This research is funded by the Federal Ministry of the Interior of Austria (2022-0.392.231).

\bibliographystyle{unsrt}

\end{document}